\documentclass[a4paper,fleqn,usenatbib]{mnras}

\usepackage{newtxtext,newtxmath}

\usepackage[T1]{fontenc}
\usepackage{ae,aecompl}

\usepackage{graphicx}	
\usepackage{amsmath}	
\usepackage{amssymb}	
\usepackage{deluxetable}

\usepackage{titlesec}
\usepackage[switch]{lineno}

\usepackage[monochrome]{xcolor}

\titleformat{\paragraph}
{\normalfont\normalsize}{\theparagraph}{1em}{}
\titlespacing*{\paragraph}
{0pt}{3.25ex plus 1ex minus .2ex}{1.5ex plus .2ex}

\title[Multiple emission regions in 3C\,84?]{KVN observations suggest multiple $\gamma$-ray emission regions in 3C\,84}

\author[J. A. Hodgson et al.]{Jeffrey A. Hodgson$^{1}$\thanks{E-mail: jhodgson@kasi.re.kr
 \dag NPP fellow},
Bindu Rani$^{2,\dag}$,
Sang-Sung Lee$^{1,3}$,
Juan Carlos Algaba$^{5,1}$,
\newauthor Motoki Kino$^{4}$,
Sascha Trippe${^5}$,
Jong-Ho Park${^5}$,
Guang-Yao Zhao$^{1}$,
Do-Young Byun$^{1}$,
\newauthor Sincheol Kang$^{1,3}$,
Jae-Young Kim$^{8,5}$,
Jeong-Sook Kim${^5}$,
Soon-Wook Kim$^{1,3}$,
\newauthor Atsushi Miyazaki$^{6}$,
Kiyoaki Wajima$^{1}$,
Junghwan Oh$^{5}$,
Dae-won Kim$^{5}$,
Mark Gurwell$^{7}$
\\
$^{1}$Korea Astronomy and Space Science Institute, 776 Daedeokdae-ro, Yuseong-gu, Daejeon, 30455, Korea \\
$^{2}$NASA/GSFC, Greenbelt, MD 20771, USA\\
$^{3}$Korea University of Science and Technology, 217 Gajeong-ro, Yuseong-gu, Daejeon 34113, Korea \\
$^{4}$Kogakuin University, Academic Support Center,
2665-1 Nakano, Hachioji, Tokyo 192-0015, Japan \\
$^{5}$Department of Physics and Astronomy, Seoul National University, 1 Gwanak-ro, Gwanak-gu, Seoul 08826, Korea \\
$^{6}$Faculty of Science and Engineering, Hosei University, 372 Kajino-cho, Koganei, Tokyo 1848584, Japan \\
$^{7}$Harvard Smithsonian Center for Astrophysics, Cambridge, MA 02138, USA \\
$^{8}$Max-Planck-Institut f\"{u}r Radioastronomie, Auf dem H\"{u}gel 69, 53121, Bonn, Germany \\ }

\date{Accepted XXX. Received YYY; in original form ZZZ}

\pubyear{2017}

\begin{document}
\label{firstpage}
\pagerange{\pageref{firstpage}--\pageref{lastpage}}
\maketitle

\begin{abstract}
3C\,84 (NGC\,1275) is a well-studied mis-aligned Active Galactic Nucleus (AGN), which has been active in $\gamma$ rays since at least 2008. We have monitored the source at four wavelengths (14\,mm, 7\,mm, 3\,mm and 2\,mm) using the Korean VLBI network (KVN) since 2013 as part of the interferometric monitoring of $\gamma$-ray bright AGN (iMOGABA) program.  3C\,84 exhibits bright radio emission both near the central supermassive black hole (SMBH) feature known as C1 and from a moving feature located to the south known as C3. Other facilities have also detected these short-term variations above a slowly rising trend at shorter wavelengths, such as in $\gamma$-ray and 1\,mm total intensity light-curves. We find that the variations in the $\gamma$ rays and 1\,mm total intensity light-curves are correlated, with the $\gamma$ rays leading and lagging the radio emission. Analysis of the 2\,mm KVN data shows that both the $\gamma$ rays and 1\,mm total intensity short-term variations are better correlated with the SMBH region than C3, likely placing the short-term variations in C1. We interpret the emission as being due to the random alignment of spatially separated emission regions. We place the slowly rising trend in C3, consistent with previous results. 
Additionally, we report that since mid-2015, a large mm-wave radio flare has been occurring in C3, with a large $\gamma$-ray flare coincident with the onset of this flare at all radio wavelengths.

\end{abstract}




\section{Introduction}

\begin{figure}
\includegraphics[width=\linewidth]{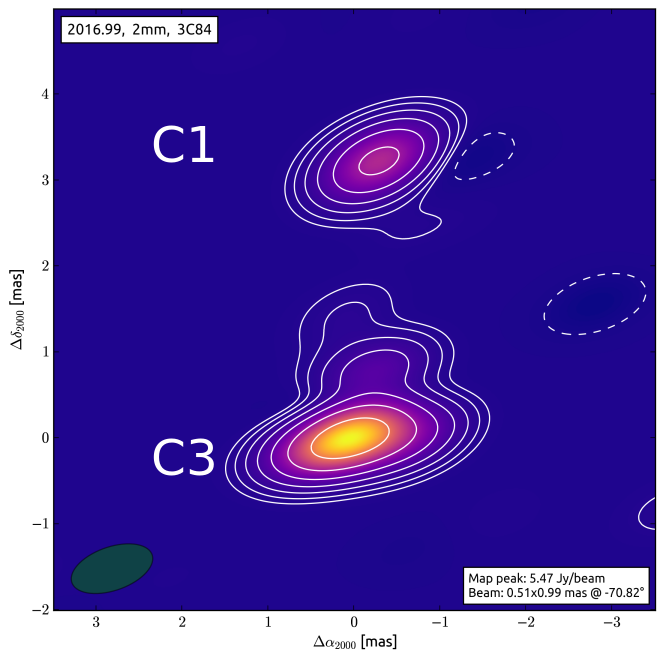}
\caption{The most recent KVN map (iMOGABA 39) of 3C\,84 at 2\,mm. C1 is thought to host the SMBH and jet launching region. C3 is a slowly moving region, having been emitted from C1 in $\sim$2003. Contours begin logarithmically at 1\% and extend to 64\% of the peak flux density in the image. \label{map} }
\end{figure}

The nearby \citep[z=0.017559, ][]{3c84_z} mis-aligned Active Galactic Nucleus (AGN) 3C\,84 (the compact radio counterpart of the Seyfert galaxy NGC 1275) is a well-studied source that has been known to be recurrently active for many decades \citep[e.g.][]{burbidge65}. It is an excellent source to investigate $\gamma$-ray physics, as it is  nearby and because it is mis-aligned; hence it is one of the few $\gamma$-ray sources where it is thought that we are able to directly observe the jet base. While the source was relatively inactive throughout the 1990s, the most recent increase in cm-radio activity began in 2005 and appeared to have been associated with the ejection of a new component in 2003 from near the base of the jet \citep{abdo09, nagai10, suzuki12}. This new component, commonly known as C3 (with the jet base being known as C1, see Fig. \ref{map} for an example of the current morphology), was emitted to the south-south-east of the jet base and is slowly moving and accelerating but also observed to be the source of the slowly increasing radio flux densities. \citet{dutson14} suggested that C3 could in fact be the base of the jet with C1 having been ejected from it; however \citet{nagai16} presented a high-resolution 7\,mm Very Long Baseline Array (VLBA) map of the source showing that the source has a morphology inconsistent with this interpretation and hence interpret C3 as being the head of a radio lobe. No correlation, however, was found with the short time scale variability of the $\gamma$ rays \citep{nagai12}. The nature of the C3 component has been broadly interpreted as likely being a radio lobe caused by the restarting of the jet interacting with the external medium and likely the source of the slowly increasing $\gamma$-ray emission \citep{nagai16}. \\

The radio structure extends further to the south, with space VLBI images showing a lobe and hot spot extending as far as approximately 20\,mas south of the central engine, which is thought to have originated during the activity in the 1950s \citep{asada06}. This radio lobe has most recently been studied by \citet{fujita16}, who showed that Bondi accretion cannot account for the observed jet power. Most recently, the source has been detected in late December 2016 and early January 2017 in very-high-energy $\gamma$ rays \citep{3c84_atel16_1, 3c84_atel16_2, 3c84_atel16_3, 3c84_atel16_4, 3c84_atel16_5}. \\


\begin{figure*}
\includegraphics[width=\linewidth]{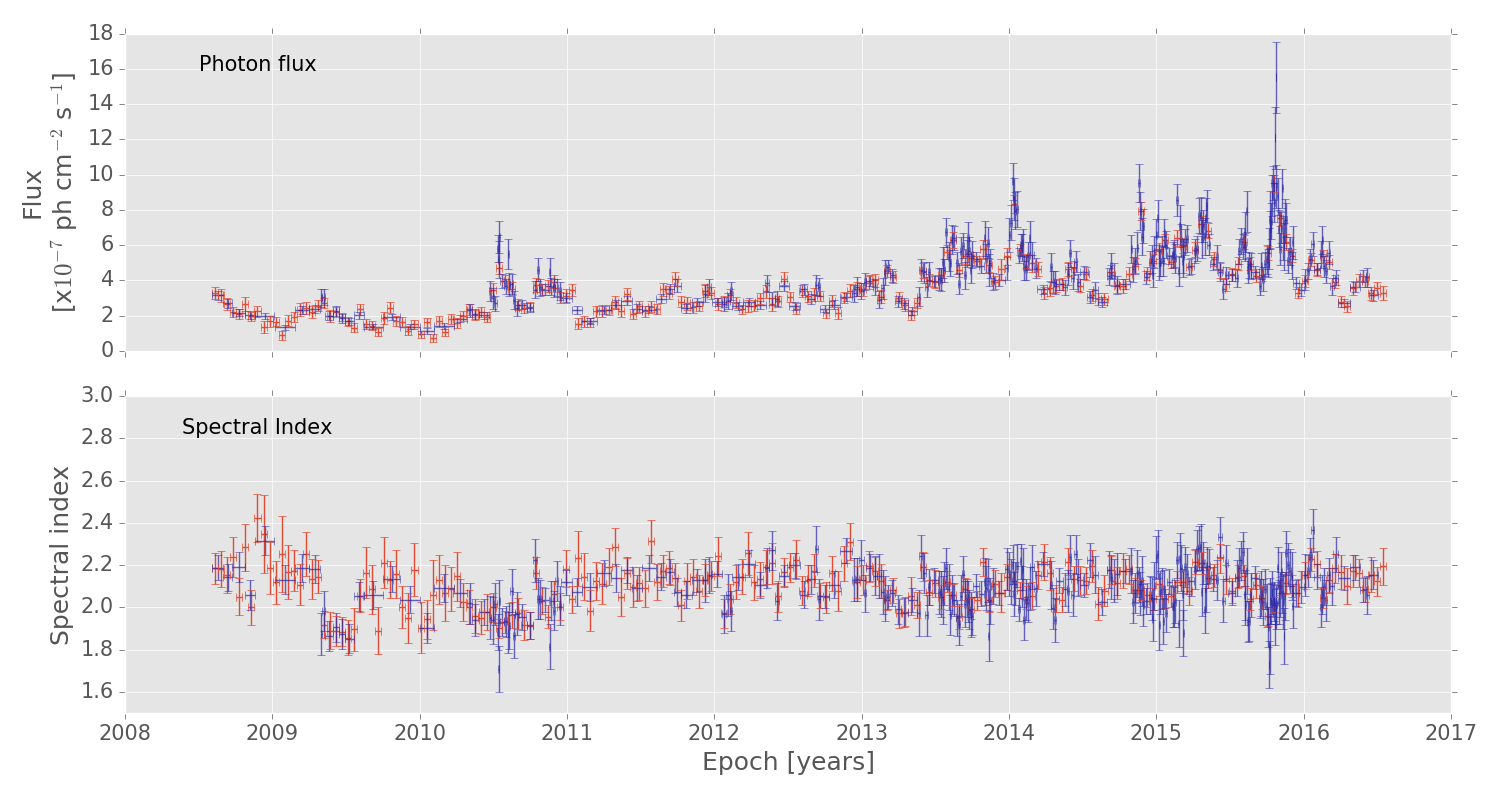}
\caption{Gamma-ray photon flux evolution (top) and spectral index (bottom). 15-day averaged photon fluxes are plotted in red and adaptively binned photon fluxes are over-plotted in purple. The average $\gamma$-ray flux has been rising since the beginning of the observations, with increased activity since $\sim$2013. 
 \label{gamma}}
\end{figure*}

\begin{figure}
\includegraphics[width=\linewidth]{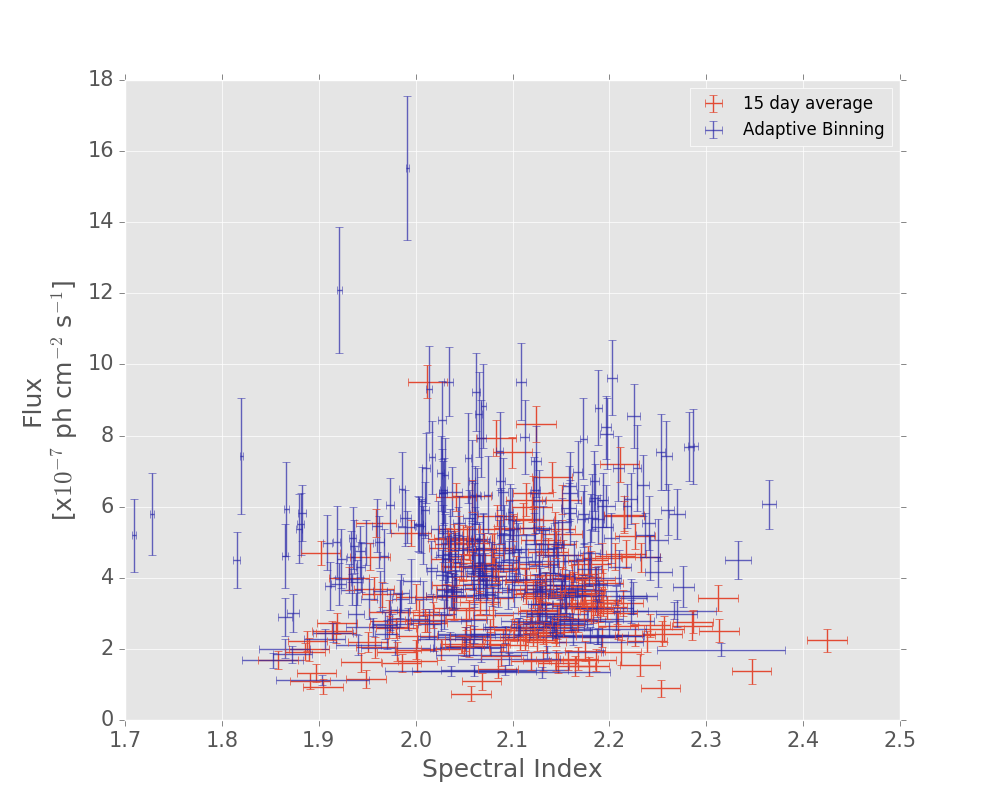}
\caption{$\gamma$-ray photon flux vs spectral index. 15-day averaged photon fluxes are plotted in red and adaptively binned photon fluxes are over-plotted in purple. \label{flux_spex}}
\end{figure}

\begin{figure}
\includegraphics[width=\linewidth]{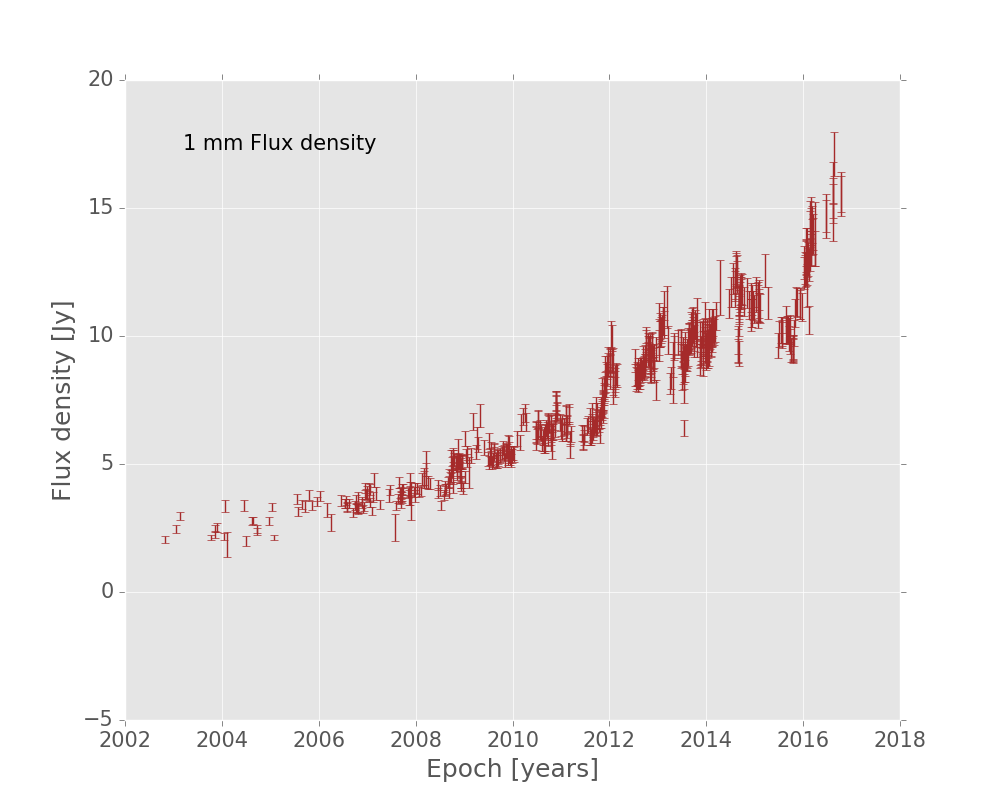}
\caption{1\,mm total intensity radio flux density evolution. Flux densities are for the entire source.  \label{sma}}
\end{figure}

\begin{table}\label{obs}
\centering
\caption{An overview of the observations performed as part of the iMOGABA program. KU: Ulsan, KY: Seoul, KS: Jeju Island. Observations that were performed as part of iMOGABA but were not successful are not shown. }
\begin{tabular}{ccc}
\hline
Code & Date& Wavelength (mm)  \\
\hline \hline
iMOGABA2 & 2013 Jan 16 & 14/7/3/2  \\
iMOGABA4 & 2013 Mar 28 & 14/7/3  \\
iMOGABA5 & 2013 Apr 11 & 14/7/3/2  \\
iMOGABA6 & 2013 May 08 & 14/7/3/2  \\
iMOGABA7 & 2013 Sep 24 & 14/7/3/2  \\
iMOGABA8 & 2013 Oct 15 & 14/7/3/2 \\
iMOGABA9 & 2013 Nov 20 & 14/7/3/2  \\
iMOGABA10 & 2013 Dec 24 & 14/7/3/2  \\
iMOGABA11 & 2014 Jan 27 & 14/7/3/2  \\
iMOGABA12 & 2014 Feb 28 & 14/7/3/2  \\
iMOGABA13 & 2014 Mar 22 & 14/7/3/2  \\
iMOGABA14 & 2014 Apr 22 & 14/7/3/2  \\
iMOGABA16 & 2014 Sep 01 & 14/7/3/2  \\
iMOGABA18 & 2014 Oct 29 & 14/7/3  \\
iMOGABA19 & 2014 Nov 28 & 14/7/3  \\
iMOGABA20 & 2014 Dec 26 & 14/7/3/2  \\
iMOGABA22 & 2015 Feb 24 & 14/7/3/2  \\
iMOGABA23 & 2015 Mar 26 & 14/7/3/2  \\
iMOGABA24 & 2015 Apr 30 & 14/7/3/2  \\
iMOGABA27 & 2015 Oct 23 & 14/7/3  \\
iMOGABA28 & 2015 Nov 30 & 14/7/3/2  \\
iMOGABA29 & 2015 Dec 28 & 14/7/3/2  \\
iMOGABA30 & 2016 Jan 13 & 14/7/3/2  \\
iMOGABA31 & 2016 Feb 11 & 14/7/3/2  \\
iMOGABA32 & 2016 Mar 01 & 14/7/3/2  \\
iMOGABA34 & 2016 Apr 25 & 14/7/3/2  \\
iMOGABA36 & 2016 Aug 23 & 14/7/3/2  \\
iMOGABA37 & 2016 Oct 18 & 14/7/3/2  \\
iMOGABA38 & 2016 Nov 27 & 14/7/3/2  \\
iMOGABA39 & 2016 Dec 28 & 14/7/3/2  \\
\hline
\end{tabular}
\end{table}

\begin{figure*}
\includegraphics[width=0.45\linewidth]{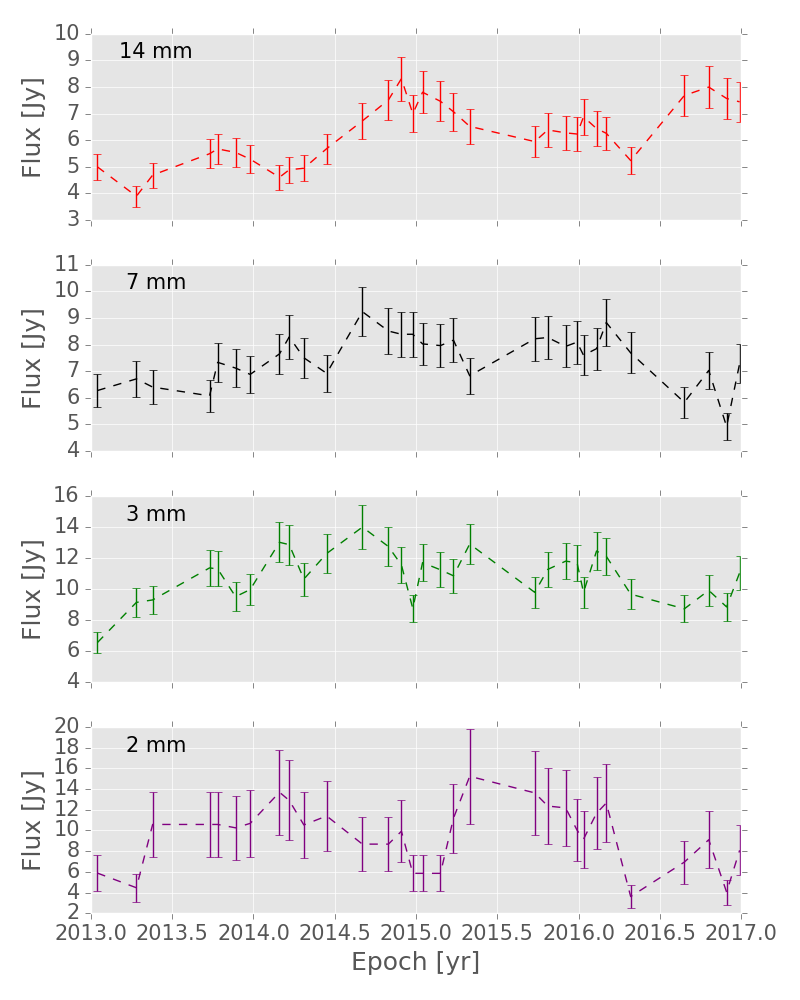}
\includegraphics[width=0.45\linewidth]{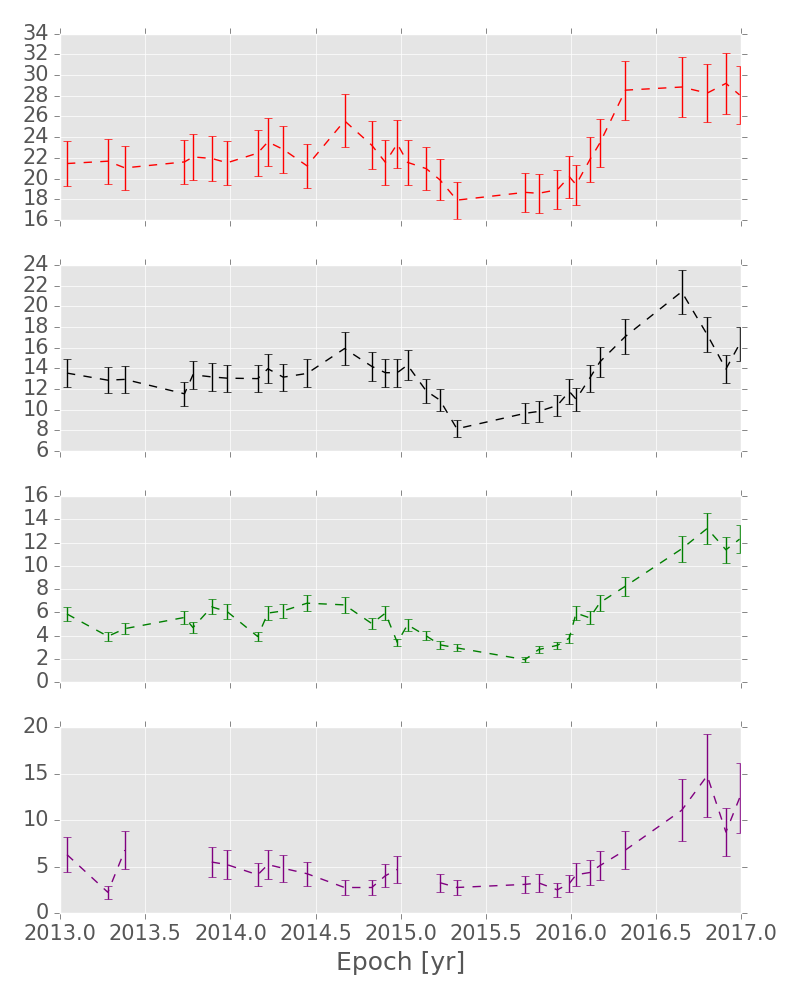}
\caption{Decomposed light curves for C1 (left) and C3 (right). Dashed lines connect the data points to guide the eye.  \label{KVN} }
\end{figure*}

\begin{figure*}
\centering
\includegraphics[width=0.95\linewidth]{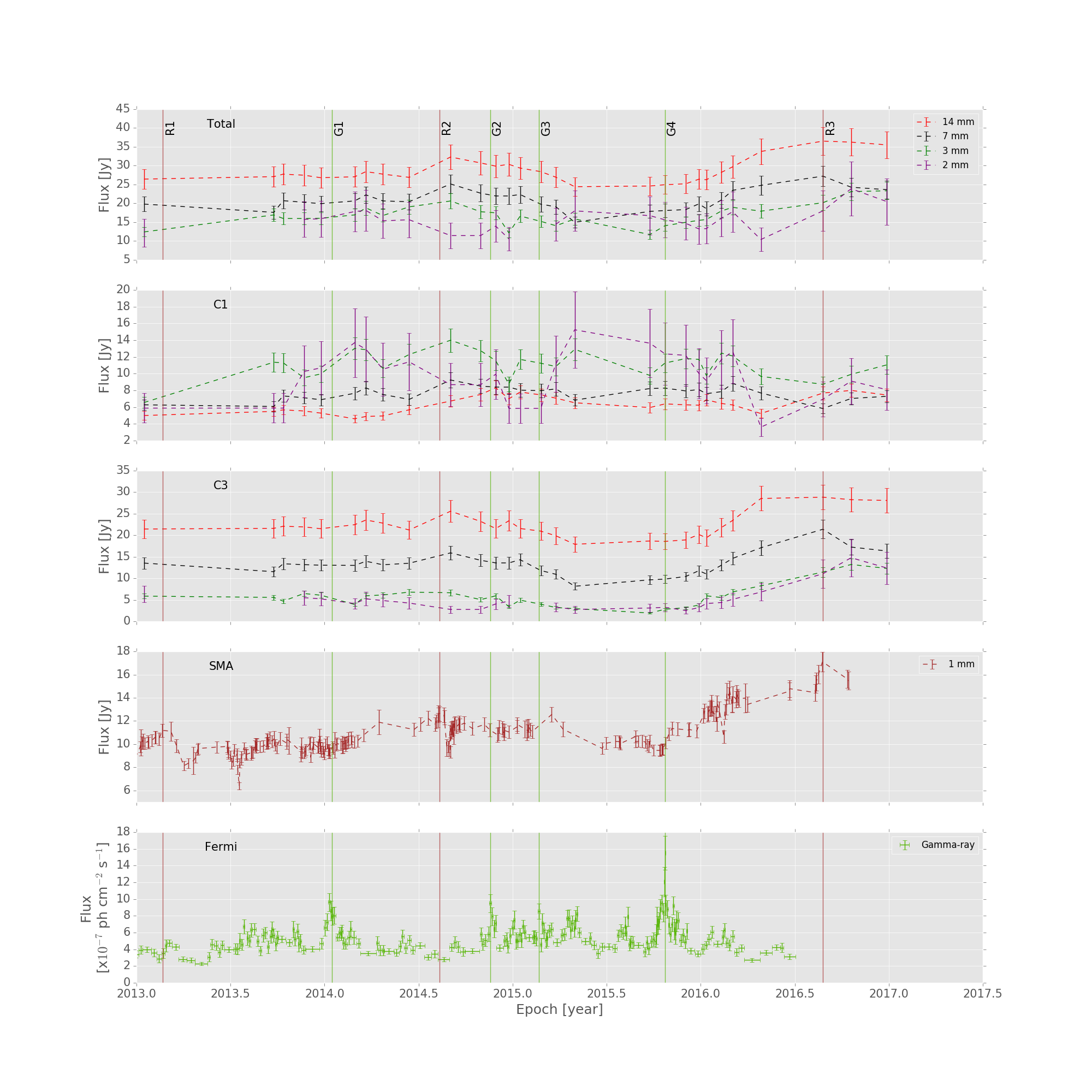}
\caption{Multi-wavelength light curves of 3C\,84. From top to bottom: total recovered flux density, decomposed light-curve for C1, decomposed light curve for C3, 1\,mm total intensity and monthly averaged $\gamma$-ray light curve. The peaks of the four most significant $\gamma$-ray flares are shown as green vertical lines. The three significant radio flares as determined from the 1\,mm total intensity light-curves (R1 - R3) are shown in  red. \label{imoLCs}}
\end{figure*}


\begin{figure*}
\centering
\includegraphics[width=\linewidth]{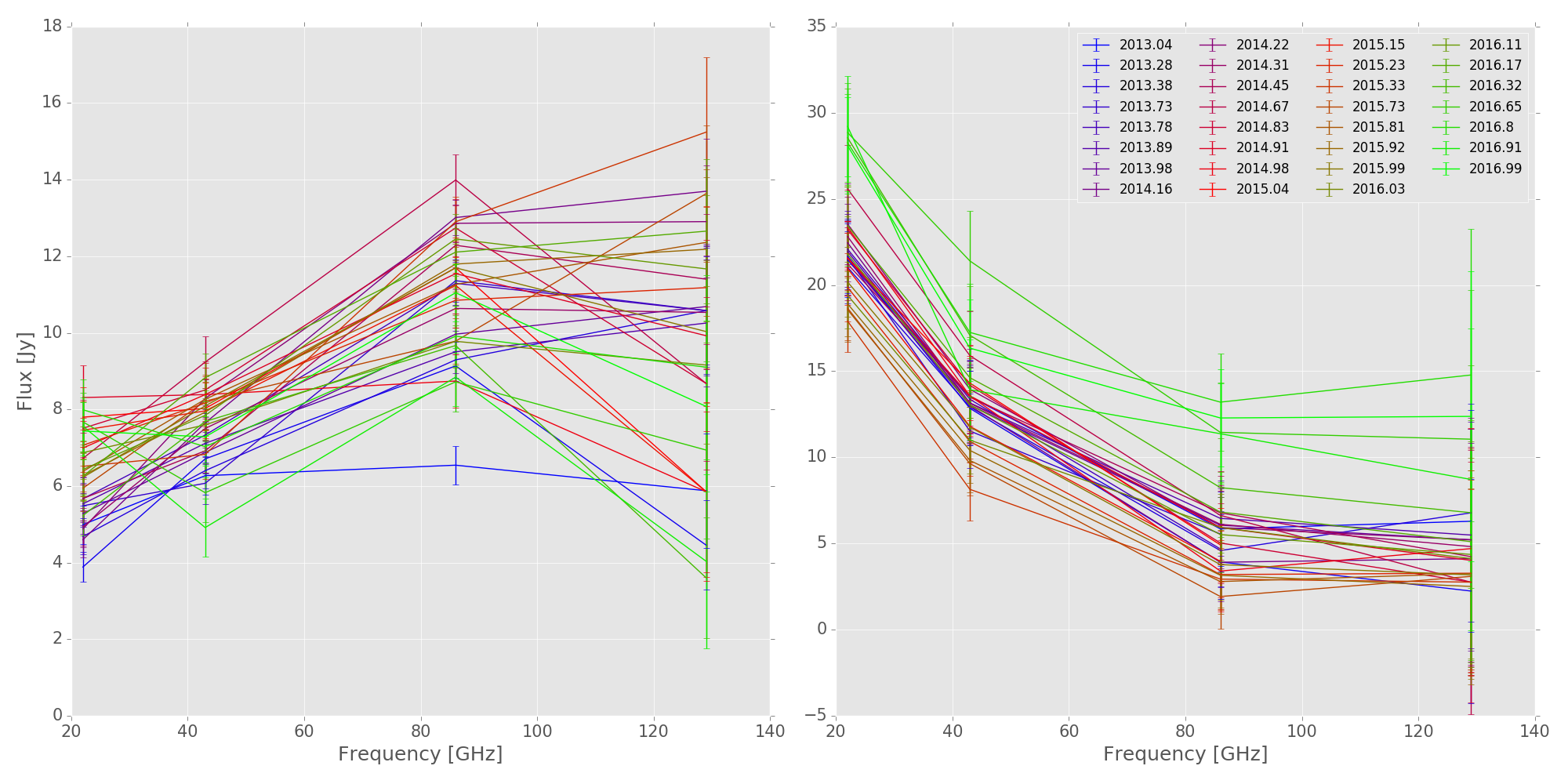}
\caption{Multi-wavelength spectra of component C1 (left) and C3 (right). The 3\,mm-2\,mm spectra in C1 is observed to be highly variable, in comparison with C3, where the 3\,mm-2\,mm spectra is observed to be approximately as variable as at longer wavelengths.  \label{spectra}}
\end{figure*}

In this paper, we present high-cadence observations of 3C\,84 using the Korean VLBI Network (KVN), which has the unique ability to observe at four wavelengths (14\,mm, 7\,mm, 3\,mm and 2\,mm) simultaneously. We combined this data with $\gamma$-ray observations from the \emph{Fermi}/LAT space telescope and with Sub-millimetre Array (SMA) 1\,mm total intensity observations. We find that some $\gamma$ rays likely originate in the C3 region while short-time scale variability is better correlated with C1 mm-wave radio emission, suggesting multiple simultaneous sites of $\gamma$-ray emission within the same source.  


In Section \ref{observations}, we present an overview of our observations. In Section \ref{results}, we present our results and analysis. In Section \ref{discussion}, we discuss our results, and in Section \ref{conclusions} we present our conclusions and our future outlook. Preliminary results have been previously reported by \citet{hodgson3c84}. In this paper we assume a standard flat $\Lambda$CDM cosmology with $H_{0}$=69.6 and $\Omega_{m}$=0.286 \citep{bennett14}. This corresponds to a linear scale of 1\,mas = 0.359\,pc at a luminosity distance of $D_{L}$=74.047\,Mpc.

\section{Observations and data reduction}\label{observations}

\subsection{Gamma-ray observations} 
Observations at GeV energies were done in a survey mode by the {\it Fermi}-LAT (Large 
Area Telescope) \citep{atwood09}. To investigate the flux and spectral variations, we used the {\it Fermi}-LAT data taken from August 2008 until October 2016. We analysed source event class photons using the standard ScienceTools (software version v10.r0.p5) and instrument response functions P8R2$\_$SOURCE$\_$V6. We chose a region of interest (ROI) of 20$^{\circ}$  radius centred at the position of the source. The selected ROI was analysed using a maximum-likelihood algorithm \citep{mattox96}. In the unbinned likelihood analysis\footnote{http://fermi.gsfc.nasa.gov/ssc/data/analysis/\\
scitools/likelihood\_tutorial.html}, we included all the sources sources from the 3FGL catalog \citep{acero15} within 20$^{\circ}$ and the recommended Galactic diffuse background ($gll\_iem\_v06.fits$) \\
and the isotropic background ($iso\_P8R2\_SOURCE\_V6\_v06.txt$) emission components \citep{Acero16}.  \\

Model parameters for the sources within 5$^{\circ}$ of the center of the ROI were set free. For the rest, we fixed the model parameters to their catalog values unless the sources were reported as being significantly variable  (variability index $\geq$72.44) in the 3FGL catalog. Model parameters for the variable sources were also left free. \\

To investigate the photon flux and index variations in the source, we generated the constant uncertainty (15\%) light curve above 100\,MeV through the adaptive binning method following \citep{lott12}. A simple power-law model  (N$_0$ E$^{-\Gamma}$, N$_0$: prefactor, and $\Gamma$: power-law index) was used to estimate the model parameters. Additionally, 2-day averaged, 15-day averaged and monthly averaged light curves were produced. The adaptively binned $\gamma$-ray photon fluxes are displayed in purple and the 15-day averaged photon fluxes are shown in red in the top panel of Fig. \ref{gamma}, with the spectral index displayed in the bottom panel. The spectral index is plotted against photon flux in Fig. \ref{flux_spex}. 



\subsection{1\,mm total intensity observations}

Total intensity radio observations at 1.3\,mm (225\,GHz) were obtained as part of the ongoing monitoring program of potential mm-wave calibrator sources at the sub-millimeter array (SMA) in Hawaii, United States \citep[see][for more details]{Gurwell07}. Scans of 3 -- 5 mins were performed with the flux density calibrated against known sources such as solar system objects. Observations were performed since October 2002 approximately weekly to monthly and are presented in Fig. \ref{sma}.

\subsection{KVN observations}\label{KVN_obs_section}


3C\,84 was observed as part of the interferometric monitoring of Gamma-ray bright AGN (iMOGABA) program that is  a Key Science Program on the KVN \citep{imo2,imo1}. Observations are performed approximately monthly (with reduced observations over the summer months), beginning in January 2013 and are ongoing. A summary of observations is given in Table \ref{obs}. The observations were performed at four wavelengths simultaneously: 14\,mm (22\,GHz), 7\,mm (43\,GHz), 3\,mm (86\,GHz) and 2\,mm (129\,GHz) \citep[see][for more details]{lee14}. Data were observed in single polarisation (left hand circular polarisation) at a recording rate of 1\,Gbps. Data were automatically reduced using the KVN Pipeline \citep{hodgson17}, which also implements the frequency phase transfer in order to improve detection rates at shorter wavelengths. See \citet{rioja11,rioja14} and \citet{imo2} for further details. \\

In order to analyse the data, maps were produced using the CLEAN algorithm within the Difmap package \citep{difmap,clean}. Circular Gaussians were fitted directly to the visibilities in order to parametrise the flux density, size and relative position angle (PA) and relative separation from a reference point. In this case, the brightest, most northern component, labelled C1, was taken as the reference point. Errors were taken as 10\% for the flux density and relative separation. A conservative error of 30\% was taken for the 2\,mm data points because of a lack of methods to conclusively determine the calibration accuracy at this observing wavelength \citep{lee14}. A more detailed discussion of the errors used is presented by \citet{hodgson17}.  \\

When comparing flux densities measured at different wavelengths, the resolution of the interferometer must be considered. At shorter wavelengths, finer structure can be resolved, hence probing different physical regions. In order to account for this effect, model-fit components at all wavelengths were therefore fixed at approximately one fifth of the beam at 14\,mm or 0.8\,mas. Hence, although finer structure could be detected at shorter wavelengths, only a simple 2-model component was fit in order to produce a like-for-like comparison. In general, if the sizes were not fixed, the best-fit sizes of the model-fit components were often close to 0.8\,mas at all bands except 2\,mm. However, the fitted sizes at 2\,mm were considerably smaller than 0.8\,mas. Fixing the size therefore increases the flux density measured for a component at 2\,mm. The combined flux density of these two components was typically within 10\% of the total CLEAN flux density at all frequencies, indicating that most flux in the source was recovered. The decomposed light-curves (LCs) are shown in Fig. \ref{KVN} and then reproduced for comparison with the other light-curves in Fig. \ref{imoLCs}. \\

\section{Results}\label{results}

\subsection{Total intensity light-curves}

\subsubsection{Gamma rays}

In Fig. \ref{gamma}, we see that since 2008, the $\gamma$-ray photon fluxes (using the adaptively binned photon fluxes as a reference) have increased over 300\%, from an average of 1.85$\pm$0.01 [x10$^{-7}$] ph cm$^{-2}$ s$^{-1}$ in 2009 to an average of 5.61$\pm$0.03 [x10$^{-7}$] ph cm$^{-2}$ s$^{-1}$ in 2015. The time range shown in Fig. \ref{gamma} is significantly different from that observed as part of the iMOGABA program. Additionally, more rapid variations are observed to be occurring with timescales on the order of weeks to months superimposed on this increasing trend. There was, however, no detection of $\gamma$ rays in 3C\,84 during the EGRET (1991 -- 2000) era, suggesting that any significant $\gamma$-ray activity has been more recent \citep{abdo09,suzuki12}. There also appears to be variability in the spectral index, particularly in the 2009 -- 2011 period. Spectral variability in later years is not as pronounced. \\

We then determined the most prominent flares. For this, we first de-trended the light-curve by performing a linear least square fit to the light-curve and subtracting that. We then found the significance by simulating the light-curves 5000 times (see Section \ref{sec:dcf} for more details). While there is considerably variability in the $\gamma$-ray light-curves, we identified the four most significant $\gamma$-ray flares, labelled G1 -- G4, which are shown in Table \ref{gamma_flares}. In Fig \ref{flux_spex}, we can see no obvious correlation between spectral index and photon fluxes in the 15-day averaged data, but potentially a softer-when-brighter trend in the adaptively binned data. To investigate this, we computed the Spearman correlation coefficient for both data sets and found no significant correlation between photon flux and spectral index in the 15-day binned data (correlation coefficient$=0.06$, $p=0.47$) and a weak but significant correlation in the 2-day binned data (correlation coefficient$=0.12$, $p=0.016$), although the trend is not significant in the adaptively binned data (correlation coefficient$=0.05$, $p=0.47$), suggesting that the softer-when-brighter trend is an effect seen in the very rapid variability data. This also suggests that important physical information may be lost when averaging the data over longer timescales. Related to this, flare durations in 3C\,84 are quite short, with even the longest flares being less than $\sim$2\,months in duration; hence the times of the flare peaks in Table \ref{gamma_flares} are reasonable approximations of the times of the flare onset. Flare durations were determined by finding by eye the local minima around the flare peak. \\



\begin{table*}
	\centering
	\caption{Significant $\gamma$-ray flares since 2013 as determined from the adaptively binned light-curve.}
	\label{gamma_flares}
	\begin{tabular}{lccccc} 
		\hline
		ID & Date & Onset & Decay & Photon flux & Spectral Index\\
		   &      &  [yr] &  [yr] & [x10$^{-7}$] ph cm$^{-2}$ s$^{-1}$ & \\
		\hline
		G1 & 2014.02 &  0.15 & 0.19 &  9.62 $\pm$ 1.28 & 2.16 $\pm$ 0.10 \\
		G2 & 2014.89 &  0.16 & 0.08 &  9.51 $\pm$ 1.26 & 2.15 $\pm$ 0.10 \\
		G3 & 2015.14 &  0.08 & 0.09 &  8.55 $\pm$ 1.17 & 2.27 $\pm$ 0.10 \\
		G4 & 2015.81 &  0.11 & 0.18 & 15.51 $\pm$ 1.27 & 1.97 $\pm$ 0.06 \\
		\hline
	\end{tabular}
\end{table*}

\subsubsection{1\,mm total intensity light-curves}

Monitoring at 1\,mm has been performed since 2002, and in Fig. \ref{sma}, we see a similar trend as for $\gamma$ rays:  rising flux densities with shorter scale fluctuations superimposed. There is a dip in the observed flux density in early 2016, that is perhaps similar to a dip in the $\gamma$ rays in early 2015, although there is no obvious correlation with equivalent flux density increases. Similar to the $\gamma$-ray light-curves,  the time-range covered in Fig. \ref{sma} is significantly different from that covered by the iMOGABA program. Also using a similar method as with the $\gamma$-ray light-curves, the 1\,mm light curve was de-trended for and simulated 5000 times for determining flare significance. The light curve is displayed for convenience also in Fig. \ref{imoLCs}. There are two significant flares (labelled R1 and R2, see Table \ref{sma_flares}), in early 2013 and mid 2014. These flares are relatively insignificant compared to a large flare seen in the 1\,mm light-curve beginning in late 2015 and peaking at 2016.65, with the peak labelled R3. The onset of this large flare is coincident with $\gamma$-ray flare G4.The durations of the flares were estimated by measuring the time between local minima on each side of the flare peak, although this is necessarily only a rough guide to the flare duration. In the case of flare R3, the flare is observed to be still ongoing. Nevertheless flares R1 and R2 are observed to be much shorter in duration than the ongoing flare R3. 


\begin{table}
	\centering
	\caption{Significant 1\,mm radio flares since 2013.}
	\label{sma_flares}
	\begin{tabular}{lccc} 
		\hline
		ID & Date & Width     & Flux density \\
		   &      &  [months] & [Jy]  \\
		\hline
		R1 & 2013.14  &  $\sim$5    & 11.19 $\pm$ 0.56 \\
        R2 & 2014.61 & $\sim$3      & 12.67 $\pm$ 0.65 \\
        R3 & 2016.65 & >14  & 17.11 $\pm$ 0.85 \\
		\hline
	\end{tabular}
\end{table}

\subsection{iMOGABA data}

The data observed with the KVN mostly had four wavelength bands  simultaneously observed, with only 4 epochs missing at 2\,mm. Unfortunately, due to summer weather and antenna maintenance constraints in Korea, there is a lack of observations during the northern summer period. These data were decomposed into two components: C1 and C3, with their decomposed light-curves shown in Fig. \ref{KVN}. The total (C1+C3) light curve is presented along with the monthly averaged $\gamma$-ray light-curves for convenience in Fig. \ref{imoLCs}. In both C1 and C3, the flux density evolution is very different at 2\,mm as compared with the lower frequencies. As seen in the 1\,mm total intensity light curve, there is a large flare detected at all frequency bands beginning in late 2015, which we associate with flare R3. Despite the conservative use of errors for the 2\,mm KVN observations, some calibration errors appear to be obvious. For example, a large increase in flux density is observed in  C1, C3 and combined flux densities in mid-2013 and also in late 2016, where a similar decrease in flux densities is observed. Since these changes occur in both components, they are likely due to calibration inaccuracies rather than being physical. We have included all available data in Fig. \ref{KVN}; however obviously bad data were removed from Fig. \ref{imoLCs} and from subsequent analysis.


\subsubsection{Component C1}

Component C1 is observed to be considerably more variable, particularly at shorter wavelengths. The lack of observations over the northern summer period means that it is difficult to conclusively identify correlations between the light-curves, but the flux density peaks at 2\,mm and 3\,mm at approximately 2014.3, approximately 4 months after flare G1.  There is also a large increase in the C1 flux density that peaks in approximately 2015.2, which is approximately 2 months after flare G3, four months after flare G2 and 14 months after flare G1. A large drop in flux density is seen in C1 early-to-mid 2016 at 2\,mm, which could be a calibration error; however as the drop is not seen also in C3, we consider it likely to be physical. \\

The flux density as a function of frequency is shown in Fig. \ref{spectra}. We find that the spectrum of C1 is highly variable, particularly at 86-129 GHz. The variability is significant and greater than our conservative 30\% error bars. It is mostly consistent with an optically thick spectrum, but is sometimes optically thin at 2\,mm. This occurs during mm-wave flares, as can be seen in Fig. \ref{imoLCs}. \\

\subsubsection{Component C3}\label{section_C3}

Before the flare R3, component C3 was also variable, but less so than C1. The flux density at 22\,GHz rises until late 2014, before decaying. This behaviour is similar at 7\,mm, but the 3\,mm and 2\,mm light-curves are observed to have a different shape. There is evidence that the flux density is rising again in 2016 at all wavelengths. There does not appear to be any obvious correlation between the $\gamma$ rays and the variability in C3 before flare R3; however the $\gamma$-ray flare G4 is coincident with the onset of the large flare. 

The spectrum of C3 has a broadly consistent spectral shape and is consistent with optically thin emission as previously reported by \citet{nagai16}. In the most recent data, during the flare R3, the spectrum of C3 has flattened, with the 7\,mm, 3\,mm and 2\,mm data becoming brighter relative to the 14\,mm data. \\
 
 We find that component C3 is slowly moving, with fitted velocities of $\mu_{14\mathrm{mm}}$ = 0.18 $\pm$ 0.02, $\mu_{7\mathrm{mm}}$ = 0.17 $\pm$ 0.02, $\mu_{3\mathrm{mm}}$ = 0.26 $\pm$ 0.04, $\mu_{2\mathrm{mm}}$ = 0.10 $\pm$ 0.08 mas year$^{-1}$. This corresponds to $\beta_{\text{app},14\mathrm{mm}}$ = 0.21 $\pm$ 0.02, $\beta_{\text{app},7\mathrm{mm}}$ = 0.20 $\pm$ 0.02, $\beta_{\text{app},3\mathrm{mm}}$ = 0.30 $\pm$ 0.03, $\beta_{\text{app},2\mathrm{mm}}$ = 0.12 $\pm$ 0.17 c. \\
 

\subsection{Correlation analysis}

In this section we analyse the correlations between the light-curves. Analysing light-curves such as these which have multiple varying components can be very difficult without having enough data to fully disentangle the components. For this reason, we have used multiple methods. In Section \ref{sec:dcf}, we use the standard discrete correlation function (DCF). We produce flux-flux plots in Section \ref{KVN_shift}. in Section \ref{sec:pearson}, we analyse using the Pearson correlation coefficient. 

\subsubsection{1\,mm total intensity-$\gamma$-ray correlations}\label{sec:dcf}

\begin{figure}
\includegraphics[width=\linewidth]{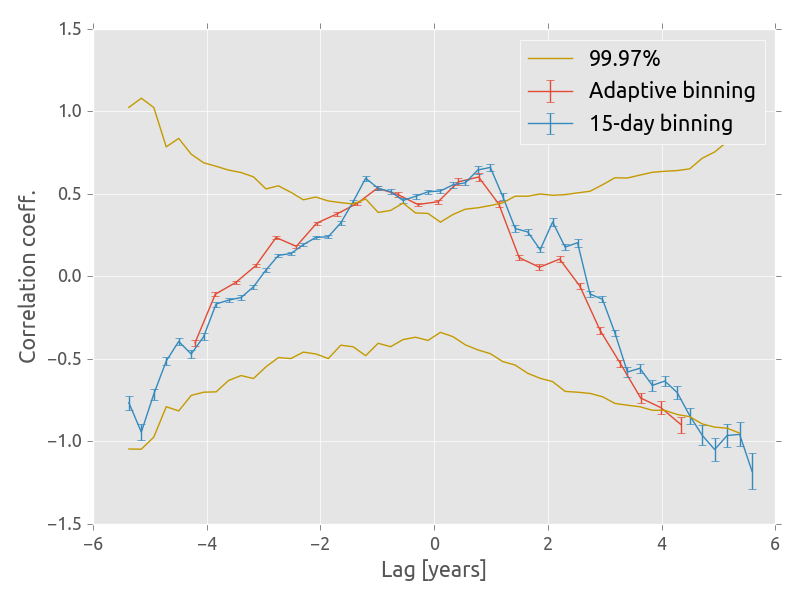}
\caption{DCF of $\gamma$ rays and 1\,mm total intensity light-curves. DCF of the adaptively binned $\gamma$-ray light-curve with 1\,mm total intensity light-curve is shown in red, 15-day binned $\gamma$-ray light-curve is in blue, and the 99.97\% significance interval is shown in gold. Two significant peaks are seen, at $-370 \pm 120$\,days and $290 \pm 155$\, days. \label{dcf} }
\end{figure}


\begin{figure}
\includegraphics[width=\linewidth]{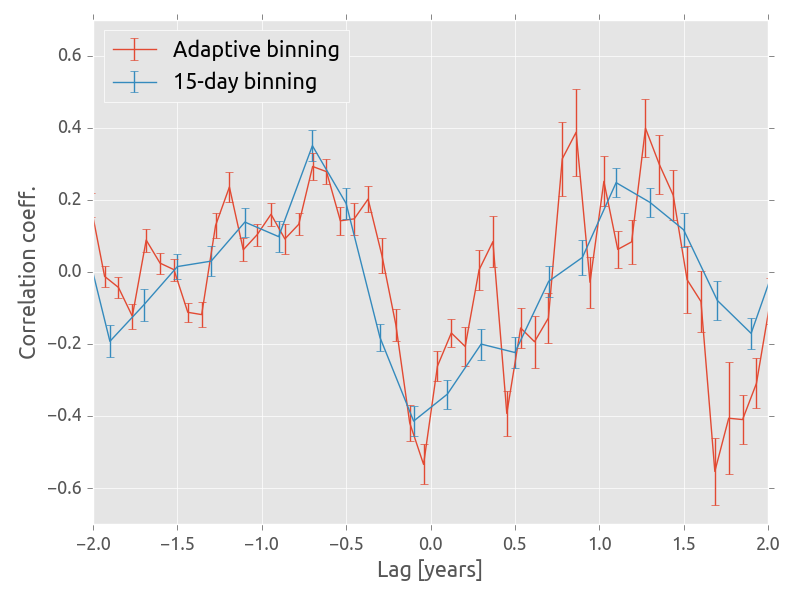}
\caption{DCF of de-trended $\gamma$ rays and 1\,mm total intensity light-curves. DCF of the adaptively binned $\gamma$-ray light-curve with the 1\,mm total intensity light-curve is shown in red; 15-day binned $\gamma$-ray light-curve is in blue. \label{dcf_detrend}} 
\end{figure}

We used the DCF \citep{dcf} method to test the apparent correlation between $\gamma$-ray and 1\,mm radio data. The 1\,mm versus adaptively binned and 15-day averaged $\gamma$-ray DCF analysis curves are shown in Fig. \ref{dcf}.  The presence of  peaks at $-370 \pm 120$\,days and $290 \pm 155$\,days indicate multiple time lags. Significance of the time lags was determined by simulating the light-curves 5000 times using DELCgen code (described by \citet{delgen}, where the code is based on \citet{emman13}). This code also takes into account the probability density function (PDF) of the given light curve. The decision to simulate the light-curves 5000 times was suggested by \citet{max14}, as lower numbers of simulations can lead to incorrect determination of significance. The first step is to determine the power spectrum density (PSD) of the data \citep{vaughan05}. The discrete Fourier transform (DFT) of the observed $\gamma$-ray data was fit using a power-law model: P(f) = $Nf^{-a}$, where $a$ is the slope and $N$ is the normalisation constant, using a least-squares fit method in log-scale. We refer to \citet{celine16} for further details of the PSD analysis. For the evenly sampled monthly averaged $\gamma$-ray data, the PSD can be well fitted with a slope of $-$(1.05$\pm$0.18). However, this method is only suitable for evenly sampled data, and the adaptive binned data are not. In this case, we first interpolated the data using a cubic spline interpolation\footnote{https://stat.ethz.ch/R-manual/R-devel/library/stats/html/splinefun.html} and then applied the PSD method described above to the interpolated data. Simulations were used in this case to test the PSD results \citep[see:][for details]{celine16}. The best fit for the adaptively binned light-curves was $-$(1.31$\pm$0.21). Finally, for the significance calculations, we simulated light curves for PSD slopes varying between $-$1.0 and $-$1.3. 


Additionally, the effects of red noise leakage and aliasing must be checked. Although the effects of red noise leakage and aliasing on determining the PSD are well understood, they are difficult to account for \citep[e.g.][]{uttley02,max14}; thus we varied the slope of the PSD by $\pm$0.3 (the typical error of the PSD on Fermi light-curves \citep{abdo10b}). The PSD slopes used to simulate data in this case were varied between $-$0.7 and $-$1.6. We found that 
the two DCF peaks are still well above the 99.97$\%$ confidence level. 


Finally, we computed the DCF for each pair of simulated light-curves and then found the percentile for each time lag. Confidence levels are computed independently using both 15-day binned and adaptively binned $\gamma$-ray data. The gold lines in Fig. 8 show a combined 99.97\% confidence level. Since the two  peaks are well above the 99.97\% confidence level, our analysis supports a significant correlation between the $\gamma$-ray photon flux and 1\,mm radio flux density variations.\\


The lags and errors corresponding to the two DCF peaks were computed by attempting different bin sizes (between 90 and 250 days, in 10-day steps) and determining the mean and standard deviation from this. Using the two DCF curves obtained using the 15-day binned and adaptively binned data, we found that the time lags corresponding to the two peaks are $-$(370$\pm$120) and 290$\pm$155 days. The positive time lag suggests that $\gamma$ rays lead radio by $\sim$300~days, while the negative time lags indicates radio flux density variations lead those at $\gamma$ rays by $\sim$400~days. \\

We see in Fig. \ref{dcf} that the $\gamma$-ray and 1 mm total intensity light-curves are highly significantly correlated. This significant correlation however could be dominated by the slow rising trend. To test the robustness of the DCF result, we removed the long-term rising 
trend from the two data sets by subtracting a linear function. The de-trended data were then used to compute a DCF; the DCF curve for the de-trended data is shown in Fig. \ref{dcf_detrend}. In general, the DCF results for the de-trended light-curves in Fig. \ref{dcf_detrend} are consistent with Fig. \ref{dcf}. This suggests that the two DCF peaks are governed by the short-term or radio flux variations. The presence of positive and negative time lags suggests
that there are most likely more than one emission region responsible for the observed variations. In the following sections, we investigate this possibilty in detail.

\subsubsection{2\,mm KVN light-curve correlations}\label{KVN_shift}

\begin{figure}
\includegraphics[width=\linewidth]{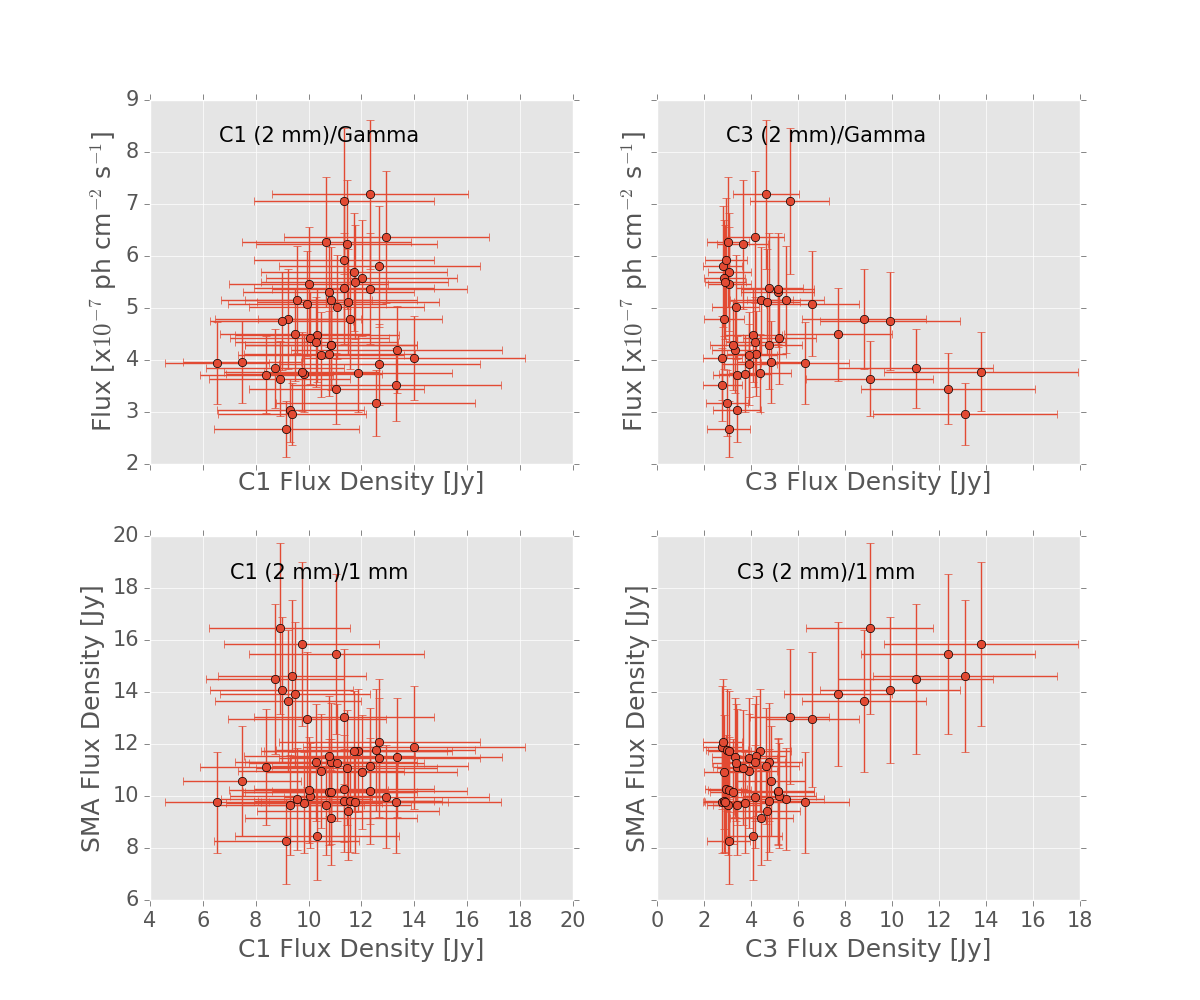}
\caption{Interpolated flux-flux plots between 2\,mm KVN light-curves and $\gamma$-ray light-curves \label{flux_flux_ALL}}
\end{figure}

\begin{figure}
\includegraphics[width=\linewidth]{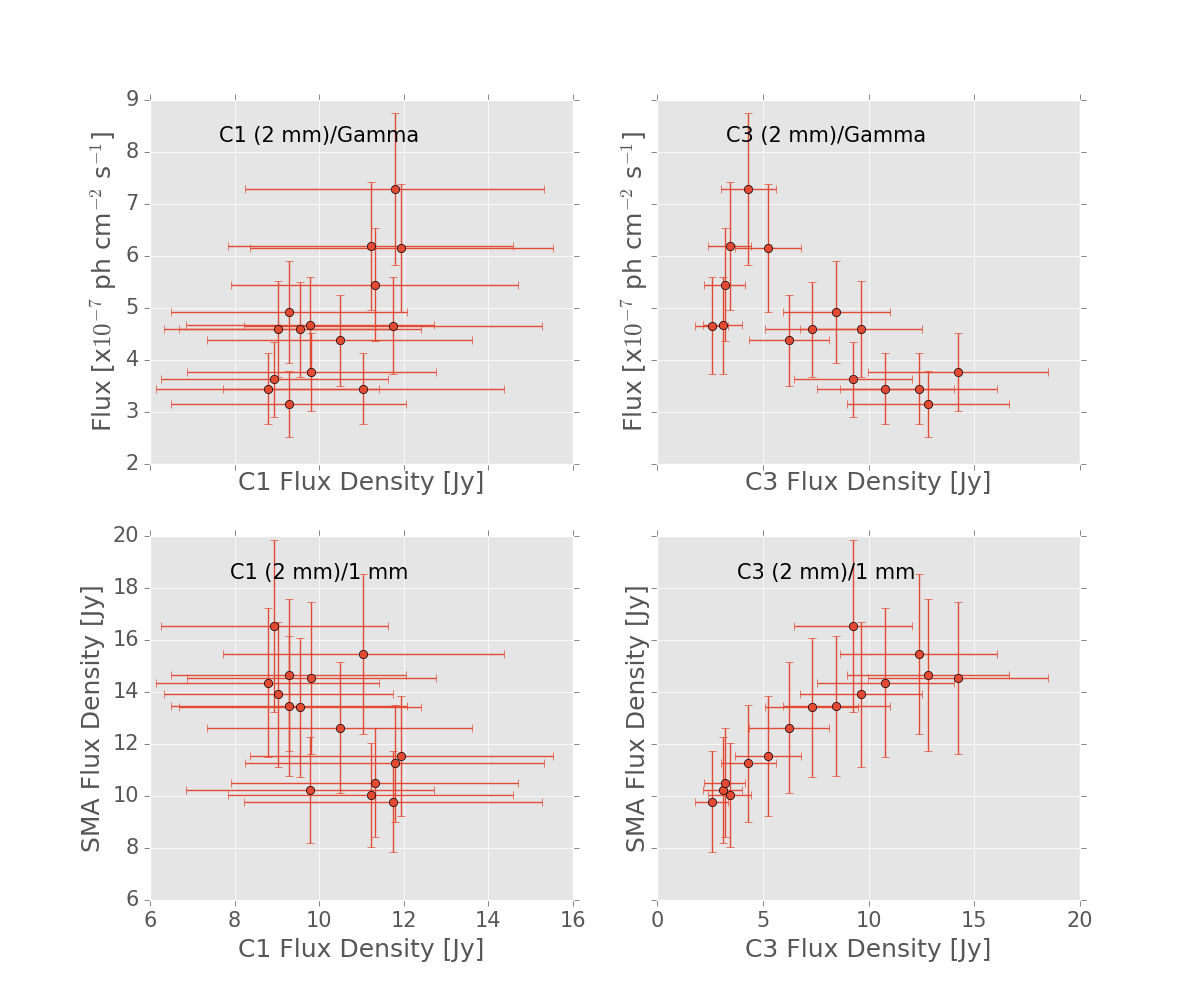}
\caption{Interpolated flux-flux plots between 2\,mm KVN light-curves and $\gamma$-ray light-curves during flare R3 \label{flux_flux_FLARE}}
\end{figure}

\begin{figure}
\includegraphics[width=\linewidth]{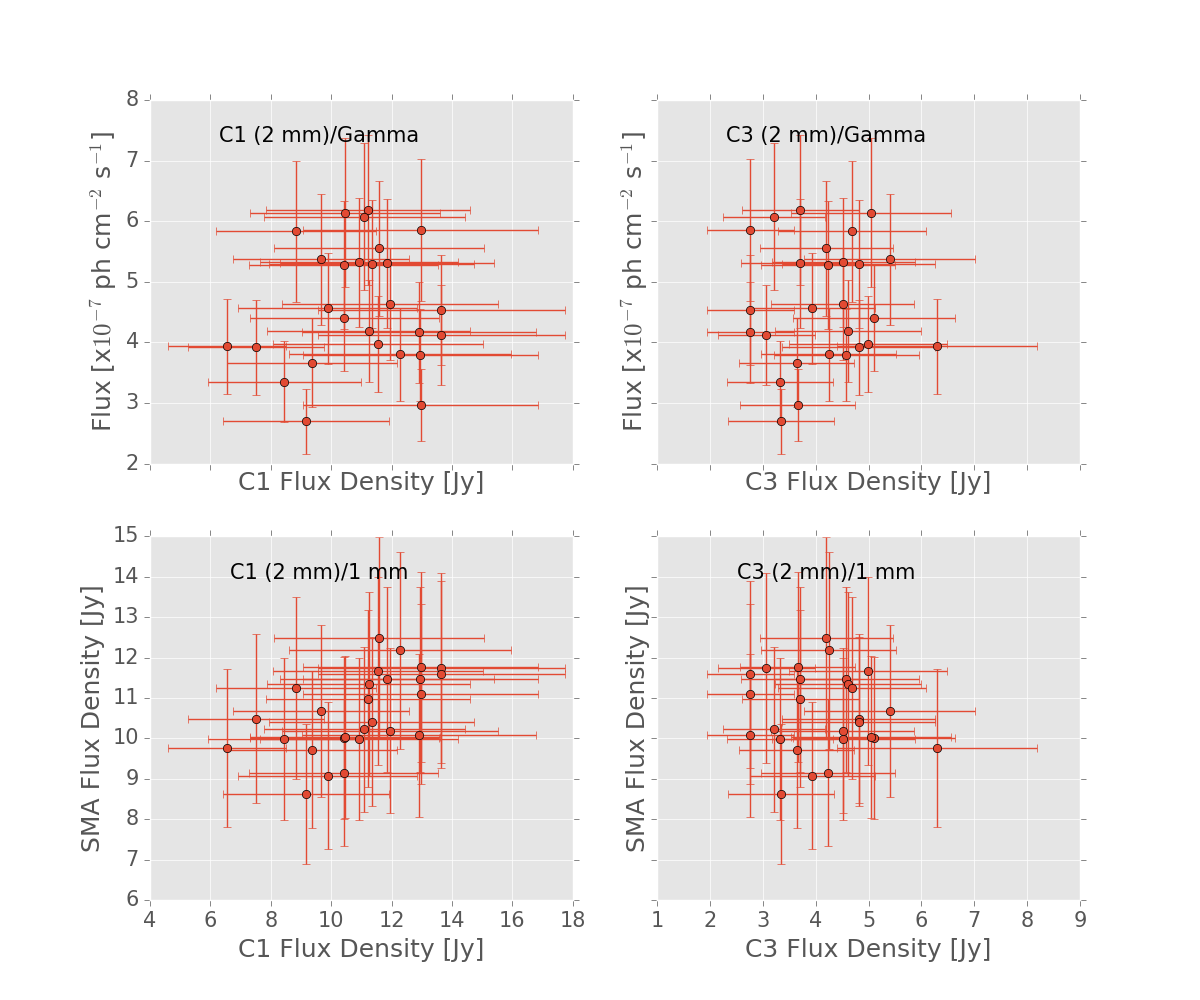}
\caption{Interpolated flux-flux plots between 2\,mm KVN light-curves and $\gamma$-ray light-curves preceding flare R3 \label{flux_flux_PREFLARE}}
\end{figure}

We then investigated if the $\gamma$ rays are better correlated with C1 or C3. Unfortunately due to the limited nature of the KVN data, we were unable to reliably perform the DCF on KVN data. However, due to the significant correlation between the 1\,mm total intensity and $\gamma$-ray light-curves, it would be expected that the 2\,mm KVN light-curves would be best correlated with the 1\,mm total intensity light curves. Hence, we re-gridded and interpolated the 2\,mm light-curves to monthly bins in order to best match the monthly $\gamma$-ray light-curves.  We then plotted flux-flux diagrams and computed the Pearson correlation coefficient for both the C1 and C3 2\,mm light-curves against the $\gamma$-ray and 1\,mm total intensity light-curves. We performed this analysis on the non de-trended data, as we have not de-trended the C1 or C3 light-curves. \\

\paragraph{Pearson analysis of all data}\label{sec:pearson}

\begin{table}
	\centering
	\caption{Pearson correlation coefficients of all data.}
	\label{pearson_all}
	\begin{tabular}{lcc} 
		\hline
		 & Coeff. & Significance \\
		\hline
        C1/Gamma & 0.13 & 0.38 \\
        C3/Gamma & $-$0.17 & 0.25 \\
        C1/1\,mm & $-$0.19 & 0.19 \\
        C3/1\,mm & 0.76 & 4.8e-10 \\
		\hline
	\end{tabular}
\end{table}

We find that there are no significant correlations between either C1 or C3 and the $\gamma$-ray light-curves when looking at all the data (Fig. \ref{flux_flux_ALL}, Table \ref{pearson_all}). However, there is a significant positive correlation between C3 and the 1\,mm total intensity light-curves, which may be dominated by flare R3. To investigate the data free from the effects of R3, we separated the data into two parts: 1) before 2015.5 (Fig. \ref{flux_flux_FLARE}, Table \ref{pearson_flare}) and 2) after 2015.5 (Fig. \ref{flux_flux_PREFLARE}, Table \ref{pearson_preflare}). \\

\paragraph{Pearson analysis during flare R3}

\begin{table}
	\centering
	\caption{Pearson correlation coefficients during flare R3.}
	\label{pearson_flare}
	\begin{tabular}{lcc} 
		\hline
		 & Coeff. & Significance \\
		\hline
        C1/Gamma & 0.58 & 0.03 \\
        C3/Gamma & $-$0.68 & 0.01 \\
        C1/1\,mm & $-$0.64 & 0.01 \\
        C3/1\,mm & 0.93 & 1.5e-6 \\
		\hline
	\end{tabular}
\end{table}

We find that during flare R3 (Table \ref{pearson_flare}), the C3 2\,mm flux density is highly positively and significantly correlated with the 1\,mm total intensity light curves. Significant anti-correlation is found between C3 and the $\gamma$-ray light-curves, which could be expected as a large $\gamma$-ray flare was detected at the onset of flare R3. An anti-correlation is found between C1 and the 1\,mm total intensity light-curves, but this is likely spurious, due to the relative dominance of the flare in C3. \\

\paragraph{Pearson analysis before flare R3}

\begin{table}
	\centering
	\caption{Pearson correlation coefficients proceeding flare R3.}
	\label{pearson_preflare}
	\begin{tabular}{lcc} 
		\hline
		 & Coeff. & Significance \\
		\hline
        C1/Gamma & 0.16 & 0.43 \\
        C3/Gamma & 0.13 & 0.52 \\
        C1/1\,mm & 0.49 & 0.01 \\
        C3/1\,mm & $-$0.25 & 0.21 \\
		\hline
	\end{tabular}
\end{table}

For the period preceding flare R3 (Table \ref{pearson_preflare}), we find no significant correlations between C1 or C3 and the $\gamma$-ray light curves, as is expected given the 8 -- 10 month offset between features in the $\gamma$-ray and 1\,mm total intensity light-curves. A significant positive correlation is however found between C1 and the 1\,mm light-curves, suggesting that the short time-scale variability seen in the 1\,mm light-curves originated in C1. No significant correlation is found between C3 and the 1\,mm light-curves. \\  

\paragraph{Pearson analysis after time-offsetting}

\begin{table}
	\centering
	\caption{Pearson correlation coefficients proceeding flare R3 with 8 month offset.}
	\label{pearson_offset}
	\begin{tabular}{lcc} 
		\hline
		 & Coeff. & Significance \\
		\hline
        C1/Gamma & 0.60 & 0.006 \\
        C3/Gamma & $-$0.32 & 0.18 \\
        C1/1\,mm & $-$0.11 & 0.63 \\
        C3/1\,mm & $-$0.46 & 0.04 \\
		\hline
	\end{tabular}
\end{table}

The results from the de-trended DCF analysis in Section \ref{sec:dcf} indicated that there is an $\sim$8 month offset between the peaks of $\gamma$-ray emission and 1\,mm total intensity radio emission. Hence, we would expect that the 2\,mm emission would also be better correlated after applying an offset of $\sim$8 months. For this reason, we offset the pre-2015.5 data by approximately 8 months (Table \ref{pearson_offset}) and recalculated the Pearson correlation coefficients. In this case, we find a significant positive correlation between C1 and the $\gamma$-ray light curves, indicating that the $\gamma$-ray light curves are better correlated with C1 than C3 before flare R3.




\section{Discussion}\label{discussion}


There appear to be three main modes of mm-wave radio emission within 3C\,84. The first is the slowly rising trend, the second is short time-scale variability and the third is the large flare beginning in approximately 2015.5. The first two modes are reflected in the $\gamma$-ray light curves, but $\gamma$-ray emission (flare G4) is only coincident with the onset of flare R3. While there is a peak in the C3 emission at all wavebands, there is no clear decay of the flare. Hence, we cannot be sure that flare R3 is still ongoing and the peaks seen are  short-term variability on top of the flare.

We note with interest that the weak softer-when-brighter trend seen in 3C\,84 in the 2\,day binned data but the lack of correlation between spectral index and photon fluxes in the 15\,day data. This is inconsistent with the harder-when-brighter trend that is typical of blazar $\gamma$-ray emission \citep{abdo10,abdo11a,rani13b}. \\


The question of the location of $\gamma$-ray emission has most recently been explored by \citet{nagai16}, who placed the longer-term $\gamma$-ray trend in C3, in agreement with \citet{dutson14}, who showed that a large emitting region is required for such slowly rising emission. As recent space VLBI images of 3C\,84 show a very compact inner structure, with emission regions on the order of hundreds of gravitational radii \citep{radio_astron_3c84}, we concur that C1 is not the source of the slowly rising trend. The location and origin of the short-timescale emission however remains elusive.  \\

In Section \ref{sec:dcf}, we found a significant correlation between the 1\,mm total intensity light-curves and the $\gamma$-ray light curves, with considerably smaller lags than that reported by \citet{dutson14}. To investigate this further, in Section \ref{KVN_shift} we shifted both the C1 and C3 light-curves by $\sim$8\,months relative to the $\gamma$-ray light curves and found that the C1 mm-wave light-curves are better correlated with the $\gamma$ rays. The DCF and KVN correlation results alone are perhaps not enough to conclusively show that the short time-scale variations originate in C1, but taken together the results are strongly suggestive that this part of the emission originates in the C1 region, which is the likely close to the immediate location of the SMBH, rather than a downstream shock. \\

Naively, if the 1\,mm and $\gamma$-ray emission delays are due to opacity effects, then we would expect the $\gamma$-ray peak to lead the 1\,mm emission \citep[e.g.][]{MG85}. Using the parameters obtained in Section \ref{section_C3}, we estimate that the $\gamma$-ray emission is $\sim$0.07\,pc upstream of the radio core, which is plausible \citep{fuhrmann14}. However this does not explain potential examples of the 1\,mm radio emission leading the $\gamma$-ray emission, the relatively insignificant correlations seen, anti-correlations and our inconsistent results compared with \citet{dutson14}. A potential explanation for these features is that the correlated emission arises from essentially random turbulent noise processes in a shocked region. \citet{marscher14} presented the Turbulent, Extreme Multi-zone (TEMZ) model, showing from numerical simulations that in this scenario, $\gamma$-ray emission would sometimes but not always be correlated with the radio emission. Thus, the observed time lags would be due to spatially separated, physically disconnected and randomly aligned (or non-aligned) emission regions, which would seem to be more consistent with our results.   \\



In Fig. \ref{spectra}, we saw that C3 is optically thin, consistent with jet emission, while the 14\,mm -- 3\,mm spectrum of the C1 region is inverted, consistent with optically thick synchrotron emission expected in the core region. This is also inconsistent with the interpretation of \citet{dutson14} that the C3 region is the base of the jet. Fig. \ref{spectra} also shows significant variability in the 3\,mm -- 2\,mm spectral index and hence more prominent variability at 2\,mm than at longer wavelengths. This perhaps explains why the short time-scale variability could not be observed in longer wavelength observations.  \\

Direct evidence of $\gamma$-ray activity so close to the central engine has been sparse, with perhaps the best example demonstrated by \citet{marti-vidal13}. To further investigate these findings, an extremely high cadence ($\sim$weekly) monitoring program is currently being undertaken on the KVN, with the results to be published in a future paper. If these results are ultimately confirmed, it may help explain why $\gamma$ rays can be highly correlated with radio emission but are only sometimes associated with component ejections in blazars. These ejections are thought to be due to travelling knots interacting with standing shocks located several parsecs downstream of the central engine, rendering the central engine largely out of view at VLBI scales \citep[e.g.][]{jor05,jor07,frank12,rani13b,hodgson17}. \\


Although the flare G4 is coincident with the onset of the flare R3 at all frequency bands, the association cannot be statistically proven and hence the detection remains tentative. The association of flares G1, G2 and G3 with C1 or C3 is not possible with the limited cadence of the KVN data. However, it does appear likely that $\gamma$-ray emission occurs in both C1 and C3, although continued observations will be required to confirm this result. \\

\section{Conclusions}\label{conclusions}

We have used simultaneous multi-wavelength observations using the KVN,  1\,mm total intensity light-curves and $\gamma$-ray light-curves to find tentative evidence for multiple sites of $\gamma$-ray emission in the mis-aligned AGN 3C\,84. There are two broad areas of emission with one likely being located near the central engine (C1) and a slowly moving feature (C3). We find that:

\begin{itemize}
  \item Short time-scale variations are observed on top of a long-term increasing trend in both the $\gamma$-ray and 1\,mm total intensity light-curves.
  \item We find a significant correlation between 1\,mm total intensity light-curves and $\gamma$-ray light-curves with the $\gamma$ rays leading the 1\,mm emission peak by approximately 8\,months. Additionally, we find a significant anti-correlation.
  \item We find significant flux density fluctuations at 2\,mm in C1.
  \item The flux variations at 2\,mm in the region of the SMBH proceeding the late 2015 flare are better correlated with the 1\,mm total intensity light-curves than C3. They are also significantly correlated with the $\gamma$-ray light curves when offset by approximately 8\,months, in agreement with the correlation analysis.   
  \item These correlations are interpreted as potentially being due to random correlations due to noise processes. 
  \item We report a large flare that is observed to originate in C3, beginning in late 2015 and possibly ongoing. The flare onset begins at approximately the same time in the 1\,mm total intensity light curves and at all KVN bands in C3.
  \item We find a large $\gamma$-ray flare coincident with the onset of of the late 2015 flare.
  \item The 2\,mm emission in C3 is highly significantly correlated with the 1\,mm total intensity emission during the large flare.
  \item We confirm the slowly increasing trend in both radio and $\gamma$ rays as originating from C3. 
  \item If these findings are confirmed with continued monitoring, it could suggest multiple locations of $\gamma$-ray emission within the jet.
  \item We detect consistent component speeds at four wavebands of $\beta_{\text{app},14\mathrm{mm}}$ = 0.21 $\pm$ 0.02, $\beta_{\text{app},7\mathrm{mm}}$ = 0.20 $\pm$ 0.02, $\beta_{\text{app},3\mathrm{mm}}$ = 0.30 $\pm$ 0.03, $\beta_{\text{app},2\mathrm{mm}}$ = 0.12 $\pm$ 0.17 c.
\end{itemize}



For follow-up work on this source, we are conducting a monitoring program of 3C\,84 using the Global mm-VLBI Array (GMVA) and a high cadence KVN monitoring program. This will allow us to explore the potential $\gamma$-ray emission regions on the scale of hundreds of gravitational radii with extremely high cadence.

\section*{Acknowledgements}

We are very grateful to the support received from staff at the KVN. The KVN is a facility operated by the Korean Astronomy and Space Science Institute. The KVN operations are supported by KREONET (Korean Research Environment Open NETwork) which is managed and operated by KISTI (Korean Institute of Science and Technology Information).
G. Zhao is supported by Korea Research Fellowship Program through the National Research Foundation of Korea(NRF) funded by the Ministry of Science, ICT and Future Planning (NRF-2015H1D3A1066561). D.-W. Kim and S. Trippe acknowledge support from the National Research Foundation of Korea (NRF) via grant NRF-2015R1D1A1A01056807. J. Park acknowledges support from the NRF via grant 2014H1A2A1018695. S. S. Lee and S. Kang were supported by the National Research Foundation of Korea (NRF) grant funded by the Korea government (MSIP) (No. NRF-2016R1C1B2006697). The Submillimeter Array is a joint project between the Smithsonian Astrophysical Observatory and the Academia Sinica Institute of Astronomy and Astrophysics and is funded by the Smithsonian Institution and the Academia Sinica. This research has made use of the NASA/IPAC Extragalactic Database (NED) which is operated by the Jet Propulsion Laboratory, California Institute of Technology, under contract with the National Aeronautics and Space Administration. This research made use of Astropy, a community-developed core Python package for Astronomy \citep{astropy}. This research made use of APLpy, an open-source plotting package for Python hosted at http://aplpy.github.com. The {\it Fermi}/LAT Collaboration acknowledges the generous support of a number of agencies and institutes that have supported the {\it Fermi}/LAT Collaboration. These include the National Aeronautics and Space Administration and the Department of Energy in the United States, the Commissariat \`a l'Energie Atomique and the Centre National de la Recherche Scientifique / Institut National de Physique Nucl\'eaire et de Physique des Particules in France, the Agenzia Spaziale Italiana and the Istituto Nazionale di Fisica Nucleare in Italy, the Ministry of Education,Culture, Sports, Science and Technology (MEXT), High Energy Accelerator Research Organization (KEK) and Japan Aerospace Exploration Agency (JAXA) in Japan, and the K.\ A.\ Wallenberg Foundation, the Swedish Research Council and the Swedish National Space Board in Sweden. Additional support for science analysis during the operations phase is gratefully acknowledged from the Istituto Nazionale di Astrofisica in Italy and the Centre National d'\'Etudes Spatiales in France. This work performed in part under DOE Contract DE-AC02-76SF00515.  This research was supported by an appointment to the NASA Postdoctoral Program at the Goddard Space Flight Center, administered by Universities Space Research Association through a contract with NASA.




\bibliographystyle{mnras}
\bibliography{Bibliography} 








\bsp	
\label{lastpage}
\end{document}